\documentclass[trackchanges,twocolumn]{aastex701}

\begin{document}


\title{Magnetic Fields in Massive Star-forming Regions (MagMaR). VI. Magnetic Field Dragging in the Filamentary High-mass Star-forming Region G35.20--0.74N due to Gravity}

\author[0000-0001-7866-2686]{Jihye Hwang}
\email[show]{astrojhwang@gmail.com}
\affil{Institute for Advanced Study, Kyushu University, Japan}
\affil{Department of Earth and Planetary Sciences, Faculty of Science, Kyushu University, Nishi-ku, Fukuoka 819-0395, Japan}

\author[0000-0002-7125-7685]{Patricio Sanhueza}
\affiliation{Department of Astronomy, School of Science, The University of Tokyo, 7-3-1 Hongo, Bunkyo, Tokyo 113-0033, Japan}
\email{}

\author[0000-0002-3829-5591]{Josep Miquel Girart}
\affiliation{Institut de Ciències de l'Espai (ICE, CSIC), Can Magrans s/n, 08193, Cerdanyola del Vallés, Catalonia, Spain}
\affiliation{Institut d’Estudis Espacials de Catalunya (IEEC), 08860 Castelldefels, Catalonia, Spain}
\email{}

\author[0000-0003-3017-4418]{Ian W. Stephens} 
\affiliation{Department of Earth, Environment, and Physics, Worcester State University, Worcester, MA 01602, USA}
\email{}

\author[0000-0003-3315-5626]{Maria T. Beltr\'an}
\affiliation{INAF-Osservatorio Astrofisico di Arcetri, Largo E. Fermi 5, I-50125 Firenze, Italy}
\email{}

\author[0000-0003-1964-970X]{Chi Yan Law}
\affiliation{Osservatorio Astrofisico di Arcetri, Largo Enrico Fermi, 5, 50125 Firenze FI, Italy}
\email{}

\author[0000-0003-2384-6589]{Qizhou Zhang}
\affiliation{Center for Astrophysics \textbar\ Harvard \& Smithsonian, 60 Garden Street, Cambridge, MA 02138, USA}
\email{}

\author[0000-0002-4774-2998]{Junhao Liu}
\affiliation{National Astronomical Observatory of Japan, 2-21-1 Osawa, Mitaka, Tokyo 181-8588, Japan}
\email{}

\author[0000-0002-3583-780X]{Paulo Cort\'es}
\affiliation{Joint ALMA Observatory, Alonso de Córdova 3107, Vitacura, Santiago, Chile}
\affiliation{National Radio Astronomy Observatory, 520 Edgemont Road, Charlottesville, VA 22903, USA}
\email{}

\author[0000-0002-8250-6827]{Fernando A. Olguin}
\affiliation{Center for Gravitational Physics, Yukawa Institute for Theoretical Physics, Kyoto University, Kitashirakawa Oiwakecho, Sakyo-ku, Kyoto 606-8502, Japan}
\affiliation{National Astronomical Observatory of Japan, National Institutes of Natural Sciences, 2-21-1 Osawa, Mitaka, Tokyo 181-8588, Japan} 
\email{}

\author[0000-0003-2777-5861]{Patrick M. Koch}
\affil{Academia Sinica Institute of Astronomy and Astrophysics, No.1, Sec. 4., Roosevelt Road, Taipei 10617, Taiwan}
\email{}

\author[0000-0001-5431-2294]{Fumitaka Nakamura}
\affiliation{National Astronomical Observatory of Japan, National Institutes of Natural Sciences, 2-21-1 Osawa, Mitaka, Tokyo 181-8588, Japan}
\affiliation{Department of Astronomical Science, SOKENDAI (The Graduate University for Advanced Studies), 2-21-1 Osawa, Mitaka, Tokyo 181-8588, Japan}
\affiliation{Department of Astronomy, Graduate School of Science, The University of Tokyo, 7-3-1 Hongo, Bunkyo-ku, Tokyo 113-0033, Japan}
\email{}

\author[0000-0002-0028-1354]{Piyali Saha}
\affiliation{Academia Sinica Institute of Astronomy and Astrophysics, No.1, Sec. 4., Roosevelt Road, Taipei 10617, Taiwan}
\affiliation{National Astronomical Observatory of Japan, 2-21-1 Osawa, Mitaka, Tokyo 181-8588, Japan}
\email{}

\author[0000-0002-6668-974X]{Jia-Wei Wang}
\affiliation{East Asian Observatory, 660 N. A`oh\={o}k\={u} Place, University Park, Hilo, HI 96720, USA}
\email{}

\author[0000-0001-5950-1932]{Fengwei~Xu}
\affiliation{Kavli Institute for Astronomy and Astrophysics, Peking University, 5 Yiheyuan Road, Haidian District, Beijing 100871, China}
\affiliation{Department of Astronomy, School of Physics, Peking University, Beijing 100871, People's Republic of China}
\email{}

\author[0000-0002-1700-090X]{Henrik Beuther} 
\affiliation{Max Planck Institute for Astronomy, Königstuhl 17, 69117 Heidelberg, Germany}
\email{}

\author[0000-0002-6752-6061]{Kaho Morii}
\affiliation{Center for Astrophysics $|$ Harvard \& Smithsonian, 60 Garden Street, Cambridge, MA 02138, USA}
\email{}

\author[0000-0001-5811-0454]{Manuel Fern\'andez L\'opez}
\affiliation{Instituto Argentino de Radioastronomía (CCT- La Plata, CONICET, CICPBA, UNLP), C.C. No. 5, 1894, Villa Elisa, Buenos Aires, Argentina}
\email{}

\author[0000-0001-9822-7817]{Wenyu Jiao}
\affiliation{Shanghai Astronomical Observatory, Chinese Academy of Sciences, 80 Nandan Road, Shanghai 200030, People's Republic of China}
\email{}

\author[0000-0003-2412-7092]{Kee-Tae Kim}
\affiliation{Korea Astronomy and Space Science Institute, 776 Daedeok-daero Yuseong-gu, Daejeon 34055, Republic of Korea}
\affil{University of Science and Technology, Korea (UST), 217 Gajeong-ro, Yuseong-gu, Daejeon 34113, Republic of Korea}
\email{}

\author[0000-0003-1275-5251]{Shanghuo Li}
\affiliation{Key Laboratory of Modern Astronomy and Astrophysics, Nanjing University, Ministry of Education, Nanjing 210023, People’s Republic of China}
\email{}

\author[0000-0003-2343-7937]{Luis A. Zapata}
\affiliation{Instituto de Radioastronomía y Astrofísica, Universidad Nacional Autónoma de México, 58090, Morelia, Michoacán, México}
\email{}

\author[0000-0002-1229-0426]{Jongsoo Kim}
\affiliation{Korea Astronomy and Space Science Institute, 776 Daedeok-daero Yuseong-gu, Daejeon 34055, Republic of Korea}
\email{}

\author[0000-0002-7497-2713]{Spandan Choudhury}
\affiliation{Korea Astronomy and Space Science Institute, 776 Daedeok-daero Yuseong-gu, Daejeon 34055, Republic of Korea}
\email{}

\author[0000-0002-8691-4588]{Yu Cheng}
\affil{National Astronomical Observatory of Japan, 2-21-1 Osawa, Mitaka, Tokyo, 181-8588, Japan}
\email{}

\author[0000-0002-8557-3582]{Kate Pattle}
\affil{Department of Physics and Astronomy, University College London, WC1E 6BT London, UK}
\email{}

\author[0000-0003-4761-6139]{Chakali Eswaraiah}
\affil{Department of Physical Sciences, Indian Institute of Science Education and Research (IISER) Mohali, Knowledge City, Sector 81, SAS Nagar 140306, Punjab, India}
\email{}

\author[0009-0007-6357-6874]{Panigrahy Sandhyarani}
\affiliation{Department of Physics, Indian Institute of Science Education and Research Tirupati, Yerpedu, Tirupati - 517619, Andhra Pradesh, India}
\email{}

\author[0000-0001-6725-0483]{L. K. Dewangan}
\affiliation{Astronomy \& Astrophysics Division, Physical Research Laboratory, Navrangpura, Ahmedabad 380009, India}
\email{}

\author[0009-0001-2896-1896]{O. R. Jadhav}
\affiliation{Astronomy \& Astrophysics Division, Physical Research Laboratory, Navrangpura, Ahmedabad 380009, India}
\affiliation{Indian Institute of Technology Gandhinagar Palaj, Gandhinagar 382355, India}
\email{}

\begin{abstract}

We investigate the magnetic field orientation and strength in the massive star-forming region G35.20-0.74N (G35), using polarized dust emission data obtained with the Atacama Large Millimeter/submillimeter Array (ALMA) as part of the Magnetic fields in Massive star-forming Regions (MagMaR) survey. The G35 region shows a filamentary structure (a length of $\sim$0.1 pc) with six bright cores located along the filament's long axis. Magnetic field strengths across the G35 region range from 0.2 to 4.4 mG with a mean value of 0.8 $\pm$ 0.4 mG. The mass-to-flux ratio ($\lambda$) varies from 0.1 to 6.0 the critical value. The highest values are found locally around cores, whereas the remains of the filament are subcritical. 
A H$^{13}$CO$^+$ (3--2) velocity gradient of 29 km s$^{-1}$ pc$^{-1}$ is evident along the filament's long axis, aligned with the magnetic field direction. At larger scales ($\sim$0.1 pc), the magnetic field lines appear roughly perpendicular to the filament's long axis, in contrast to the smaller-scale structure ($\sim$0.003 pc) traced by ALMA. 
The magnetic field lines could be dragged along the filament as a result of the gas motion induced by the gravitational potential of the filament. Six cores in the filament have similar spacings between 0.02--0.04 pc. The initial filament fragmentation could have produced a core spacing of 0.06 pc, following filament fragmentation theory, and the current core spacing is the result of cores comoving with the gas along the filament. This core migration could occur in a few 10$^4$ years, consistent with high-mass star formation time scales.
\end{abstract}


\section{Introduction} \label{sec:introduction}

Magnetic fields can play an important role in star formation processes by regulating the formation of filaments and their fragmentation into cores. In addition, magnetic fields support these dense and cold cores, the progenitors of stars, against the gravitational collapse (e.g., \citealt{Maury2022, Pattle2023} and references in the reviews).
Polarized dust emission observations can reveal the magnetic field orientation in star-forming regions. In general, dust polarization emission has been believed to be caused by the alignment of spinning dust grains, the minor axes of which are  parallel to the field lines, by radiative torques \citep[RATs;][]{Lazarian2007}. 
At cloud scales of $\sim$ 4--10 pc (1.5$^\circ$ at distances of 140--400 pc), magnetic field orientations and density structures are correlated. Although magnetic field lines have been found to be preferentially perpendicular to the filamentary structures above the hydrogen column density of 10$^{21.7}$ cm$^{-2}$, they are parallel to the structures with lower densities \citep{Planck2016b}. \citet{Pillai2020} also show that this alignment of magnetic field orientations changes as densities increase in sub-pc scales. The parallel alignment of magnetic fields with lower-density structures may help guide material flows into denser regions, ultimately affecting the conditions under which core fragmentation occurs.
However, there are only a few studies that discuss the role of the magnetic field in core fragmentation. The relative importance between magnetic field, turbulence, and gravity can determine the type of core fragmentation \citep[e.g.,][]{Tang2019,Wu2024, Hwang2025, Lee2025}. 

Both the magnetic field strength and its orientation have been used to discuss the importance of the field in the star-forming process. Magnetic field strength in the plane of the sky is estimated using the Davis-Chandrasekhar-Fermi method \citep{Davis1951, ChanFer1953}. The method assumes a uniform magnetic field orientation which is distorted by non-thermal gas motions.

The Magnetic fields in Massive star-forming Regions (MagMaR) survey aims to study the role of the magnetic field in the formation of cores embedded in high mass star-forming regions.
The MagMaR results have shown diverse magnetic field morphologies and the relative importance between gravity, turbulence, and magnetic fields in several high-mass star-forming regions \citep[e.g.,][]{Fernandez2021, Cortes2021, Sanhueza2021, Cortes2024, Saha2024, Zapata2024, Sanhueza2025}.

One of the MagMaR targets, G35.20-0.74N (hereafter G35), is an active massive star-forming region located at a parallax distance of 2.19$^{+0.24}_{-0.20}$ kpc \citep{Zhang2009}. G35 contains a bright infrared source IRAS 18556 + 0136 with high bolometric luminosity ($>$ 10$^4$ $L_\odot$), a molecular outflow, radio continuum sources, and masers \citep[e.g.,][]{Gibb2003, Birks2006, Sanchez2013, Beltran2016, Wang2023}. {\it Spitzer} observations reveal a mid-IR nebula with a ``butterfly'' shape that has a direction similar to the outflow \citep{Sanchez2013}. In the center of the nebula, the integral filament with a length of $\sim$ 0.6 pc has been traced in NH$_3$ emission using the $Jansky$ Very Large Array \citep[JVLA;][]{Wang2023}. Atacama Large Millimeter/submillimeter Array (ALMA) observations in Band 7 resolved a filament with a length of $\sim$0.2 pc containing six dense cores in the central region of the NH$_3$ integral-shaped filament \citep{Sanchez2013}. The six dense cores are separated with similar projected distances. Some studies suggest that the aligned core fragmentation can be regulated by the magnetic field \citep{Tang2019, Liu2024}.
The magnetic field in the G35 filament with a length of $\sim$0.2 pc has been studied using dust polarization data from the Submillimeter Array (SMA) at 1.3 mm with a beam size of $\sim$1$''$ \citep{Qiu2013, Zhang2014}. They showed that field lines are aligned along a northern part of the filament containing four dense cores. They suggested that those field lines are affected by the gas rotation. The magnetic field lines are perpendicular to the southern part of the filament, which has been interpreted as a part of the hourglass pattern. 
Our MagMaR observations have revealed magnetic field structures in G35 with a higher resolution of 0\farcs3 compared to the previous SMA data. Although the overall morphology is consistent with that seen in the SMA observations, our new observations cover detailed magnetic field structures that were not detected.

In this work, we present polarized dust emission and H$^{13}$CO$^+$ ($J$ = 3 -- 2) line observations of the G35 region obtained by ALMA as part of the MagMaR survey. Observations and data reduction are described in Section \ref{sec:obs}. In Section \ref{sec:Res}, we present magnetic field orientations along the filamentary structure and provide a map of the estimated magnetic field strengths. To discuss the relative importance of the magnetic field and gravity, we estimate the mass-to-flux ratio in Section \ref{sec:Disc}. We investigate the role of the magnetic field in the core fragmentation within the filament based on the magnetic field orientations, strengths, and velocity gradients obtained from the H$^{13}$CO$^+$ line in Section \ref{sec:Disc}. We summarize our results and conclusions in Section \ref{sec:Sum}.

\section{Observations and Data reduction} \label{sec:obs}
\subsection{ALMA Data}
The dust polarization observations of G35 were carried out using ALMA Band 6 (242.51--261.283 GHz) on 2018 September 25 as part of the MagMaR survey (Project IDs: 2017.1.00101.S and 2018.1.00105.S; PI: Patricio Sanhueza). A total of 45 antennas with diameters of 12 m were used, covering baselines of 15 to 1400 m, resulting in an angular resolution of $\sim$0\farcs3 and a maximum recoverable size of $\sim$4\farcs5. The correlator setup contains five spectral windows: three centered at 244.447, 245.446, and 257.446 GHz for full polarized dust continuum emission observations with a spectral resolution of 1.953 MHz and a total bandwidth of 1.875 GHz each, and two centered at 260.199 and 261.166 GHz for detecting H$^{13}$CO$^+$ ($J$ = 3 -- 2) and HN$^{13}$C ($J$ = 3 -- 2) spectral lines with a spectral resolution of 488.281 kHz ($\sim$0.56 km s$^{-1}$) and a total bandwidth of 234.38 MHz each. Because  H$^{13}$CO$^+$ data shows a larger structure than HN$^{13}$C, we used H$^{13}$CO$^+$ data for this study. The noise level for H$^{13}$CO$^+$ is 1.8 mJy beam$^{-1}$ per 0.28 km s$^{-1}$ channel.  

We followed the procedures described in \citet{Olguin2021} to subtract channels with line emission from the continuum (Stoke $I$) image. We conducted three self-calibration processes in phase of Stokes $I$ with a final solution interval of 10 s.
The solution of self-calibration was also applied to the spectral cubes, which were cleaned with Briggs weighting and a robust parameter equal to 1 using the automatic masking procedure {\it yclean} presented in \cite{Contreras2018}. 

Stokes $I$, $Q$, and $U$ images were independently cleaned using the Common Astronomy Software Applications package (CASA; \citealt{CASA2022}) task \textit{tclean}, with Briggs weighting and a robust parameter of 1. The final Stokes $I$, $Q$, and $U$ images have a synthesized beam of 0\farcs30 $\times$ 0\farcs27 with a position angle of -55.9 degrees. The pixel size of the images is 0\farcs05. 
The polarization angle ($\theta_\mathrm{obs}$) and its uncertainties ($\sigma_{\theta_\mathrm{obs}}$) are estimated as follows:
\begin{equation}
\theta_\mathrm{obs} = 0.5\tan^{-1}(U/Q)
\label{eq:qu}
\end{equation}
The uncertainties of Stokes $I$ ($\delta I$), $Q$, and $U$ are 233 $\mu$Jy beam$^{-1}$, 22 $\mu$Jy beam$^{-1}$, and 22 $\mu$Jy beam$^{-1}$, respectively. These values were obtained as the root mean square in the Stokes $I$, $Q$, and $U$ emission-free regions. 
The polarized intensity image was debiased following \cite{Vaillancourt2006}. 
More details about the data reduction can be found in previous MagMaR papers \citep[e.g.,][]{Sanhueza2021}

\subsection{SOFIA and JVLA data}

To ascertain the large-scale magnetic field of G35, we used ancillary Band D (154 $\micron$) data from the Stratospheric Observatory For Infrared Astronomy (SOFIA) observatory using the HAWC+ polarimeter \citep{Dowell2010, Harper2018}. Data were taken from the SOFIA program 09\_0016 (PI: I. Stephens), which was a SOFIA survey program to observe the large-scale magnetic fields around several MagMaR sources. Calibrated polarization products were downloaded directly from the SOFIA archive, which includes debiased polarization maps. Band D observations have a spatial resolution of 13\farcs6 \citep{Harper2018}, but the pipeline applies a smoothing that makes the effective resolution 14\farcs0. Observations were in Scan mode, which used on-the-fly mapping. Both G35.13 and G35.20N were observed as part of this program, but we only focus on G35.20N (called G35 in this paper).

We also used the kinematic temperature obtained by \citet{Wang2023} to constrain the distribution of the temperature in G35. 
\citet{Wang2023} presented NH$_3$ observations obtained using the $Jansky$ Very Large Array (JVLA). The details of the observations and the data reduction are described by \citet{Wang2023}. They  derived the kinetic temperature by fitting the NH$_3$ (1,1) to (7,7) inversion transition lines, which were obtained with a resolution of 2\farcs9 $\times$ 1\farcs9. We used the kinetic temperature they obtained for our analysis.

\section{Results}\label{sec:Res}

\subsection{Magnetic Field Orientation}

\begin{figure*}[htb!]
\epsscale{0.8}
\plotone{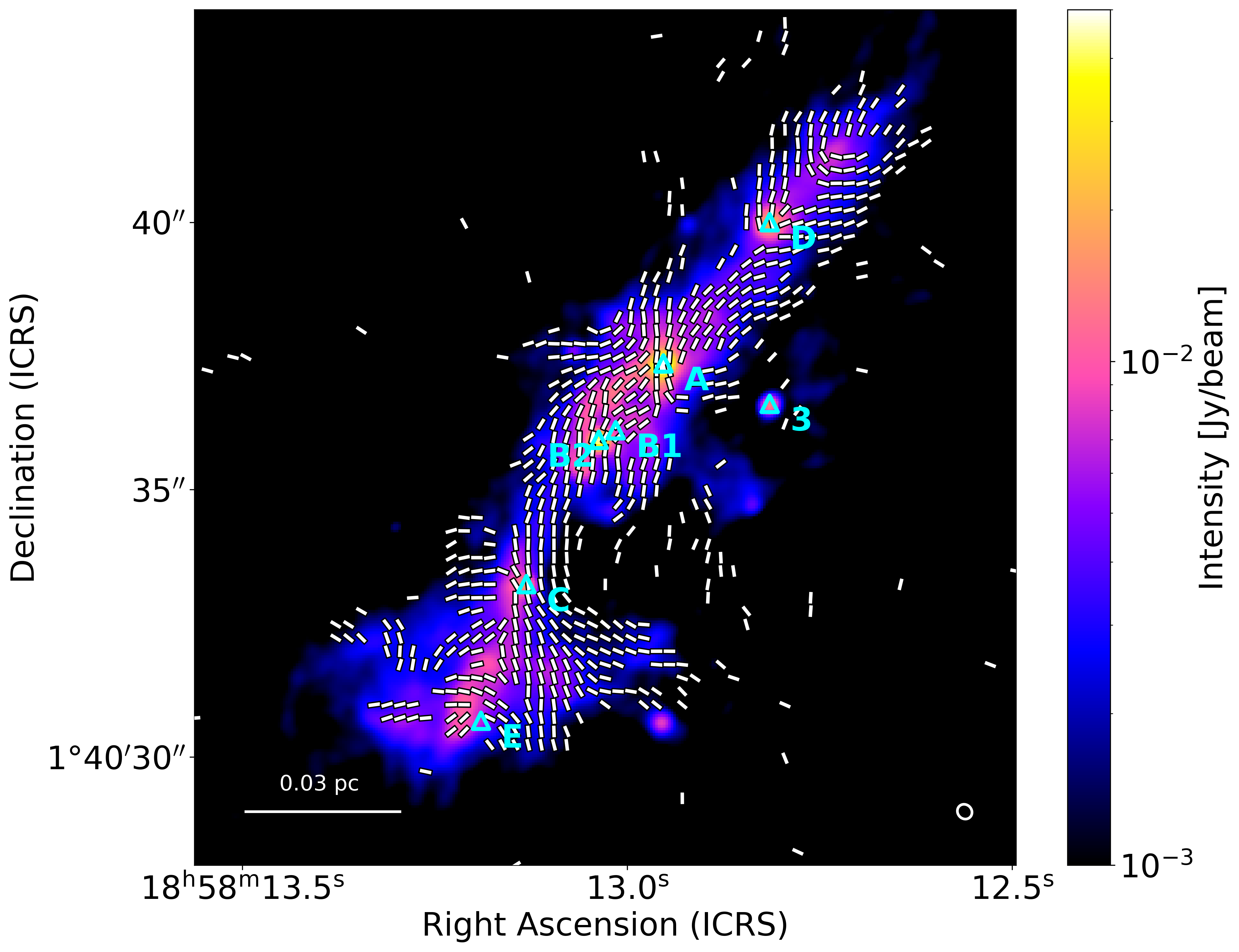}
\caption{Map of the magnetic field orientation obtained from dust polarization observations using ALMA in G35. The background image is the intensity (Stokes $I$) at Band 6 frequency, $\sim250$ GHz. The white segments show the magnetic field orientatio,n which is the polarization angle rotated by 90 degrees, and is plotted roughly per independent beam. The triangles are the positions of cores found in previous ALMA dust continuum observations with a higher angular resolution  dthan ours \citep{Zhang2022} The core names are labeled based on previous studies \citep{Zhang2022, Sanchez2013}. The physical scale and beam size are shown in the bottom left and right corners, respectively. 
\label{fig:magori}}
\end{figure*}

The polarization segments are rotated 90 degrees to show the magnetic-field orientation in G35 (Figure \ref{fig:magori}). We assume that the minor axes of the dust grains are aligned along the magnetic field lines by RATs. All segments in the figure are scaled to the same length for uniformity.
We only show segments with a spacing of 5 pixels (0\farcs25) which is similar to the ALMA beam size to clarify the orientation of magnetic field lines. The overall magnetic field orientation is aligned along the major axis of the filamentary structure traced by dust continuum emission (Stokes I).

\begin{figure*}[htb!]
\epsscale{1.0}
\plotone{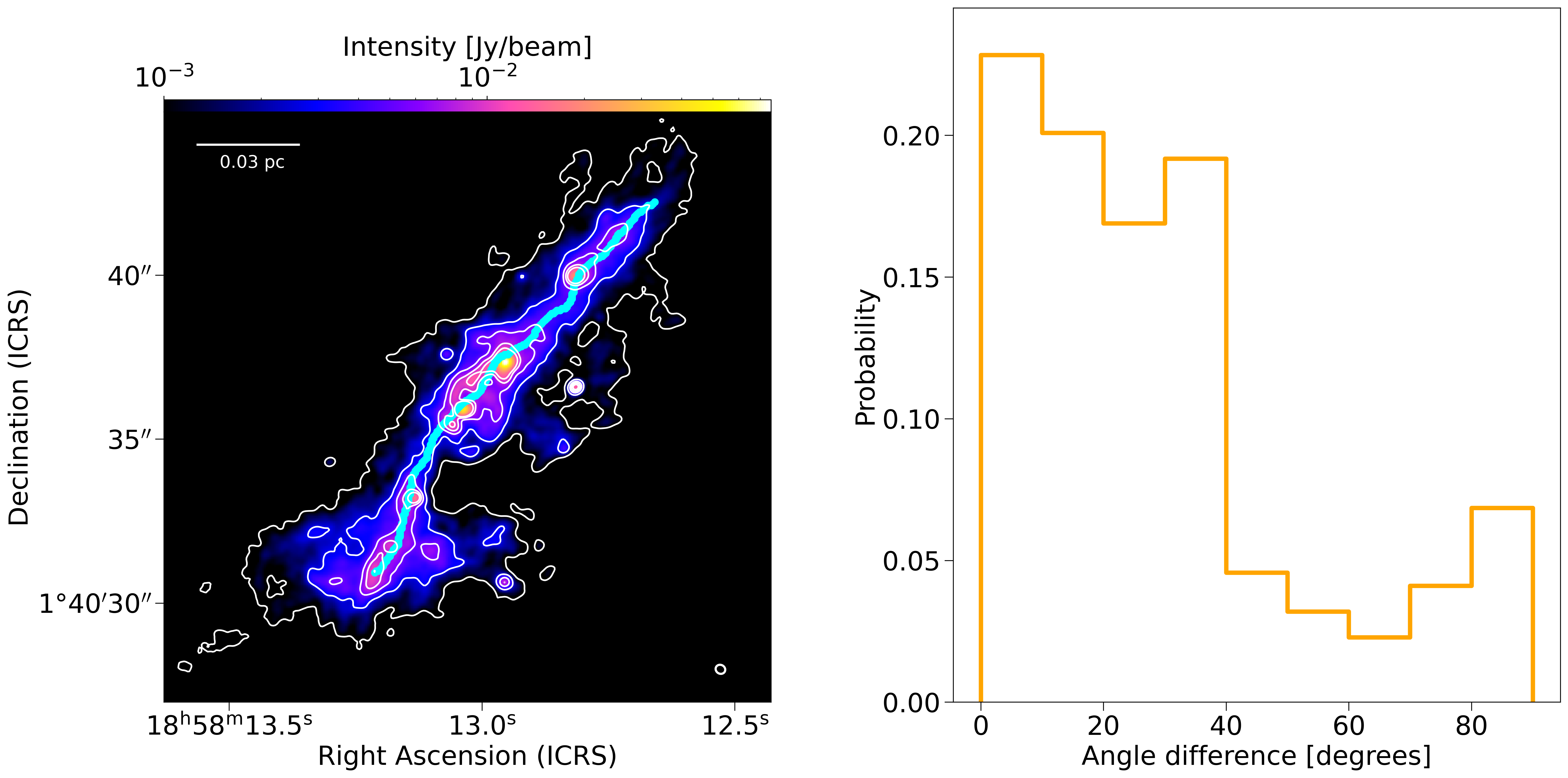}
\caption{(Left) The skeleton of the filament in G35. Background image is the same as in Figure \ref{fig:magori}. White contours show the flux density of G35 at 3, 10, 20, 30, and 40 $\times$ $\delta I$ levels. (Right) Angle difference between the main skeleton and magnetic field orientations.
\label{fig:fila}}
\end{figure*}

To show the relative alignment of the magnetic field lines and the filament, we estimate the angle difference between the skeleton of the filament and the magnetic field orientation (Figure \ref{fig:fila}). We find filamentary skeleton structures using the FilFinder algorithm \citep{Koch2015}, in which skeletons are defined using the Medial Axis Transform (MAT; \citealt{Blum1967}). The first step of FilFinder is to make a mask for the target region. Then, it finds the largest circles that can fit inside the mask without crossing the edges. By linking the centers of these circles, it extracts the skeleton of the structure. 
The black line in Figure \ref{fig:fila} shows the skeleton in G35 found by the FilFinder algorithm\footnote{We obtained the skeleton using the parameters which are minimum threshold of 11$\delta I$, minimum mask size of three beam size, and minimum length of 0.03 pc.}. 
The orientation of the skeleton at a given pixel on the black line was determined by computing the mean angle formed between the pixel and its neighboring skeleton pixels within one beam size. After calculating the orientations of all skeleton pixels, we estimate the angle difference between the magnetic field orientation and the local skeleton orientation at each pixel. The mean and median values of the angle difference are 29° and 24°, respectively (Figure \ref{fig:fila}), implying that the magnetic field and the skeleton are relatively well aligned.

\subsection{Polarization Angle Dispersion}

We estimate the polarization angle dispersion ($\sigma_\theta$) following the method introduced by \citet{Pattle2017, Hwang2021, Hwang2022}. In the DCF method, magnetic field lines are assumed as a uniform direction, but they can be curved because of the effects of gravity, outflows, or gas motions. Magnetic field lines can be assumed to have a uniform orientation in a smaller box than the curvature of the field lines, so \citet{Hwang2021} obtained a map of the polarization angle dispersion by moving the box over a star-forming region. 
They estimated the radii of curvature of the polarization segments over the region using Equation (6) in \citet{Koch2012}. We apply the same method to calculate the radii of magnetic field curvature in G35. Two polarization segments can be tangents of a circle whose radius is the same as that of the curvature. The radii of curvature are estimated from the mean values of four pairs of polarization segments, which were positioned between a given segment and its right, left, upper, and lower counterparts. Each pair for calculation of radii of curvature is separated by six pixels, comparable with a beam size. When the radii of curvature are smaller than the size of the box, the dispersion of the polarization angle could be overestimated \citep{Hwang2021}. We exclude pixels having smaller radii of curvature than the box size measuring the polarization angle dispersion.

\begin{figure*}[htb!]
\epsscale{1.2}
\plotone{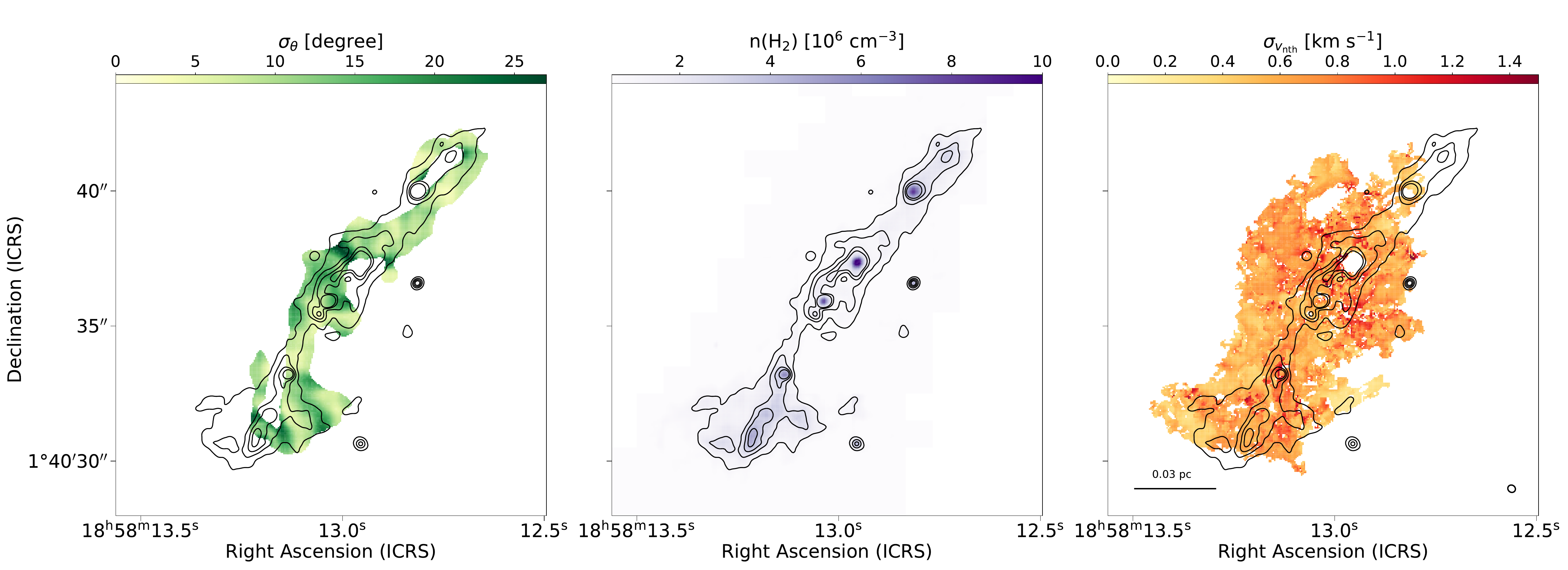}
\caption{The maps of polarization angle dispersion (Left), volume density (Middle), and non-thermal velocity dispersion of the non-thermal component of H$^{13}$CO$^+$ (Right) in G35. Black contours show the flux density of G35 at 10, 20, 30, and 40 $\times$ $\delta I$ levels. The region within 10 $\times$ $\delta I$ level is comparable to the filament with the width of 0.015 estimated in Section \ref{sec:vol} along the black skeleton shown in Figure \ref{fig:fila}. 
\label{fig:poldisp}}
\end{figure*}

We determine a box size of 18 pixels $\times$ 18 pixels (0\farcs9 × 0\farcs9), which contains approximately nine synthesized beams (corresponding to 6 pixels $\times$ 6 pixels per beam). \citet{Hwang2021} showed that the dispersion of the polarization angle depends on the ratio between the box size and the radius of curvature. If this ratio is larger than 1, the angle dispersions obtained with 3 $\times$ 3, 5 $\times$ 5, and 7 $\times$ 7 boxes are consistent (see Appendix B in \citealt{Hwang2021}).
However, if pixels with small radii of curvature are included, the polarization angle dispersion can be overestimated due to rapid changes in polarization angle over short distances.
Based on their results, we exclude pixels with a radius of curvature smaller than 18 pixels when calculating the polarization angle dispersion. We calculate polarization angle dispersion as the standard deviation of polarization angles in the box using circular statistics when the number of pixels having radii of curvature larger than 18 pixels is greater than 60\% of the total pixels within the box. The calculated dispersion value is assigned to the ninth pixel along both the Right Ascension (RA) and Declination (Dec) directions, which approximately corresponds to the center of the box. By moving the box across the map, we make a map of the polarization angle dispersion in G35 (left panel of Figure \ref{fig:poldisp}), which ranges from 3 to 27 degrees, with a mean value of 11 degrees. Using $24\times24$ and $30\times30$ pixels increases the mean value by $\sim$20\% relative to 11 degrees (to $\sim$13.2 degrees), and this additional uncertainty is negligible compared to the uncertainty in the volume density.

\subsection{Volume Density}\label{sec:vol}

We first estimate the column density of molecular hydrogen,  $N$(H$_2$), in G35 using the blackbody equation,
\begin{equation}
N(\text{H}_2) =I_\nu/(\mu  m_{\text{H}}\kappa(\nu) B_\nu(T)), 
\label{eq:col}
\end{equation}
where $I_\nu$ is the intensity at a frequency of $\nu$, $\mu$ is the mean molecular weight per particle which is 2.8 assuming 90\% and 10\% of the total number of particles to be hydrogen and helium, respectively, $m_{\text{H}}$ is the mass of hydrogen atom, $\kappa(\nu)=\kappa_{\nu_0}(\nu/\nu_0)^\beta$ is the dust opacity in which we assumed a dust opacity $\kappa_{\nu_0}$ of 1.1 cm$^2$ g$^{-1}$ at $\nu_0$ of 1.2 mm \citep{Ossenkopf1994}, the spectral index $\beta$ is 2, the gas-to-dust mass ratio is 100, and $B_\nu(T)$ is the Planck equation at dust temperature $T$. The dust temperature is assumed to be equal to the kinetic temperature obtained by \citet{Wang2023}, which is reasonable at the high density region we analyze \citep[e.g.,][]{Goldsmith2001, Krumholz2022}.
The beam size of the ammonia data is 2\farcs9 $\times$ 1\farcs9, which is larger than the box size of 0\farcs9 $\times$ 0\farcs9. However, there is no other temperature distribution in the overall G35 structure with a resolution comparable to our ALMA observations. We regridded the map of kinetic temperature to our ALMA flux density map, such that a uniform kinetic temperature value is assigned to all ALMA pixels within a single VLA pixel.

We assume that the depth of the filament is equal to its width. The width of the filamentary structure is estimated to be 0.015 pc, derived from the FWHM of a Gaussian fit to the averaged intensity profile taken perpendicular to the skeleton orientations presented in Section \ref{fig:magori}. Details of this profile analysis are provided in Appendix \ref{sec:fwhm}. We assume a cuboid volume of 0\farcs05 $\times$ 0\farcs05 $\times$ 1\farcs4 (0.0005 pc $\times$ 0.0005 pc $\times$ 0.015 pc) for each pixel. By dividing the H$_2$ column density map by the assumed depth of 0.015 pc per pixel, we obtain the H$_2$ volume density map (middle panel of Figure \ref{fig:poldisp}).
Within the lowest contour level shown in the figure, the H$_2$ volume densities, $n$(H$_2$), range from (1.6 $\pm$ 0.6) $\times$ 10$^5$ to (1.2 $\pm$ 0.5) $\times$ 10$^7$ cm$^{-3}$, with a mean value of (8.8 $\pm$ 4.6) $\times$ 10$^5$ cm$^{-3}$.
The uncertainties in the H$_2$ volume density are derived via error propagation based on Equation (\ref{eq:col}). The uncertainty on temperature is adopted from the values derived from \citet{Wang2023} by fitting multiple ammonia transitions. The fractional uncertainties in dust opacity and the gas-to-dust ratio are adopted as 23\% and 28\%, respectively, following \citet{Sanhueza2017,Sanhueza2019}. The uncertainty of flux calibration in ALMA Band 6 is approximately 10\%, according to the ALMA Technical Handbook\footnote{\url{https://almascience.nrao.edu/proposing/technical-handbook}}. The dominant sources of uncertainty in the column density arise from the dust opacity and gas-to-dust ratio.
The uncertainty in depth is less well constrained. If the filament was considered to be a cylindrical structure, the depth would decrease with increasing distance from the filament skeleton. This would result in unrealistically small volumes at the side boundary of the filament. Therefore, for simplicity and consistency, we adopt a uniform depth of 0.015 pc.

\subsection{Velocity Dispersion}\label{sec:veldisp}

We estimate the velocity dispersion of the gas using the H$^{13}$CO$^+$ $J=3-2$ emission line. We use the Python toolkit $FUNStools$\footnote{\url{https://github.com/radioshiny/funstools}} to decompose multiple Gaussian components. After decomposition of the components, we applied a friends-of-friends (fof) algorithm to find velocity-coherent groups of Gaussian components. We arbitrarily selected a pixel located between cores B and C with a simple Gaussian profile as the starting point for the algorithm. 
When the velocity and amplitude of the Gaussian components in neighboring pixels are within $\pm$0.05 km s$^{-1}$ and $\pm$5\% of the amplitude of the seed component, we connect these components with the seed. 
If multiple components satisfy these criteria in a neighboring pixel, the component with the lowest velocity difference from the seed value is chosen. After performing these processes in all neighboring pixels, the seed components are changed to the neighboring component chosen in the previous step, and the aforementioned process is repeated. This algorithm is stopped if no neighboring components satisfy the selection criteria. Then, we choose a new seed component among the remaining components that are not selected as part of the group. Using this iterative process, we can find coherent groups of velocity in G35. To find a velocity-coherent component covering the diffuse region of the filament between dense cores, we choose the first seed component at a pixel having one single component. After applying the fof algorithm, we obtained a velocity coherent group that covers most of the region of G35. We calculated the non-thermal component of the gas in the group. The velocity coherent group cannot cover the northern and central core regions (the right panel of Figure \ref{fig:poldisp}). The northern regions have different velocity structures, and the central region contains a hot molecular core in which the molecular lines have complex non-Gaussian features. 

The non-thermal component of the velocity dispersion $\sigma_{v_\mathrm{nth}}$ is estimated by subtracting the thermal component, 
\begin{equation}
\sigma_{v_\mathrm{nth}} =\sqrt{ \sigma_{v_{\text{obs}}}^2 - \frac{ k_{\rm B} T_k}{m_{\text{H}^{13}\text{CO}^+}}},
\label{eq:vel}
\end{equation}
where $\sigma_{v_{\text{obs}}}$ is the velocity dispersion obtained from the fit of the observed H$^{13}$CO$^+$ spectral line, $T_k$ is the kinetic temperature which is taken here to be that obtained by NH$_3$, and $m_{\text{H}^{13}\text{CO}^+}$ is the mass of the H$^{13}$CO$^+$ molecule of 30 atomic unit mass. The right panel of Figure \ref{fig:poldisp} shows the map of the non-thermal velocity component derived from  H$^{13}$CO$^+$. The velocity dispersion of the non-thermal gas component ranges from 0.1 to 4.0 km s$^{-1}$, with mean and standard deviation values of 0.6 $\pm$ 0.2 km s$^{-1}$ within the lowest contour of Figure \ref{fig:poldisp}. 


\subsection{Magnetic Field Strengths} \label{sec:mag}

We obtained a map of magnetic field strengths using the DCF method \citep{Davis1951, ChanFer1953}. \citet{Crutcher2004} parameterized the method using the polarization angle dispersion, the volume density, and the nonthermal gas velocity dispersion as given in the following formula, 
 \begin{equation} 
B_{\text{pos}} = Q\sqrt{4\pi\rho}\frac{\sigma_{v_\mathrm{nth}}}{\sigma_\theta} \approx 5.2\sqrt{n(\text{H}_2)}\frac{\Delta V}{\sigma_\theta},
\label{eq:cf}
 \end{equation}
where $B_{\text{pos}}$ is the magnetic field strength in the plane of the sky in $\mu$G, $Q$ is a correction factor according to the line-of-sight and beam integration effects, $\rho$ = $\mu m_\mathrm{H} n(\mathrm{H}_2)$ is the mass density and $\Delta V = \sqrt{8 \ln 2}\sigma_{v_\mathrm{nth}}$ is the full width at half maximum (FWHM) of the non-thermal gas component in km s$^{-1}$. We adopted the correction factor $Q$ of 0.28 suggested by \citet{Liu2022} in the case of a gas structure with $n(\mathrm{H}_2) = 10^4-10^6$ cm$^{-3}$ and a size of $l = 0.2-1$ pc based on numerical simulations, where $l$ is the size of a cloud. Although the simulation criteria are slightly different from our observed values ($l$ = 0.1 pc and $n(\mathrm{H}_2) = 10^5-10^7$ cm$^{-3}$), the $Q$ value estimated by \citet{Liu2022} has conditions more similar to the observations than the criteria from \citet{Ostriker2001} of $l = 8$ and $n(\mathrm{H}_2) = 10^2$ cm$^{-3}$, from which the standard value of $Q$ = 0.5 was determined. In addition, only 12 pixels in our map have a volume density larger than 10$^7$ cm$^{3}$.
 
\begin{figure*}[htb!]
\epsscale{0.8}
\plotone{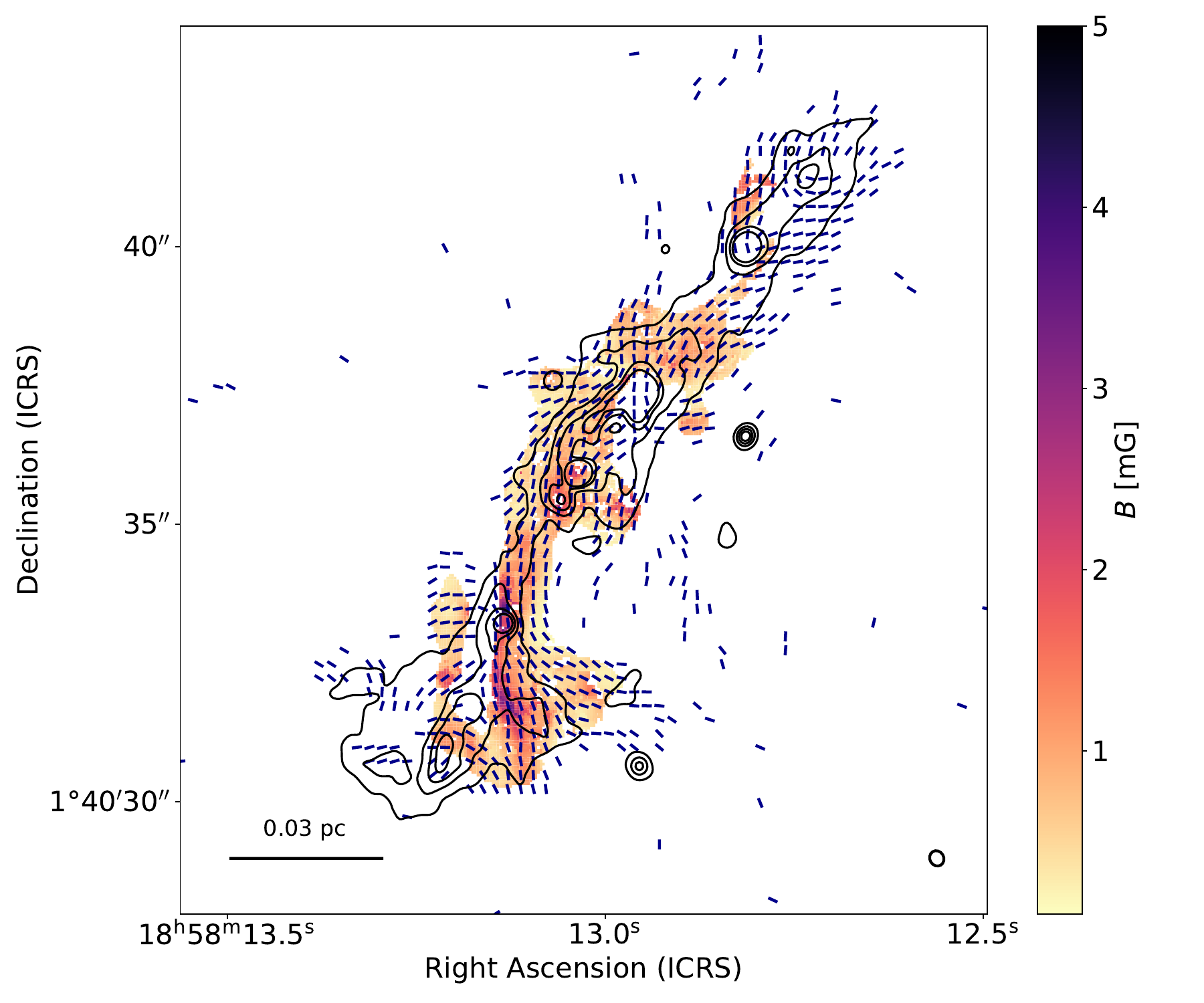}
\caption{The map of magnetic field strength in G35. Black contours are the same as shown in Figure \ref{fig:poldisp}. Blue segments represent magnetic field orientations obtained using ALMA, which are the same as those shown in Figure \ref{fig:magori}.
\label{fig:mag}}
\end{figure*}

By substituting the polarization angle dispersions, number density, and velocity dispersions in each pixel into the method, we obtained the magnetic field strength map (Figure \ref{fig:mag}). 
Within the lowest contour level of the figure, the magnetic field strength ranges from 0.2 to 4.4 mG with a mean value of 0.8 $\pm$ 0.4 mG. The uncertainty on the magnetic field strength is calculated through error propagation of equation (\ref{eq:cf}). 

\section{Discussion}\label{sec:Disc}

\subsection{Mass-to-flux Ratio} \label{sec:mtob}

To investigate the relative importance of gravity and the magnetic field, we estimated the mass-to-flux ratio $(M/\Phi)_{\rm obs}$ \citep{Mouschovias1976, Crutcher2004}. The observed mass-to-flux ratio is scaled by a critical value which is estimated by assuming a magnetized disk, $(M/\Phi)_{\rm crit}=1/2\pi G^{1/2}$ \citep{Nakano1978}, where $G$ is the gravitational constant. This critical value is 1.3 times lower than that obtained by assuming an isothermal filament \citep{Tomisaka2014}. The mass-to-flux ratio function in units of critical value \citep{Crutcher2004} is

\begin{equation}
\lambda = 7.6 \times10^{-21} \frac{N(\text{H}_2)}{B},\label{eq:mass_flux_ratio}      
\end{equation}
\noindent
where $B$ is the magnetic field strength in three dimensions (3D) in units of $\mu$G. A value of $\lambda$ less than unity indicates that the star-forming region is \textit{magnetically subcritical}, meaning the magnetic field is strong enough to resist gravitational collapse. Conversely, a value of $\lambda$ greater than unity implies that the region is \textit{magnetically supercritical}, where the magnetic field is no longer sufficient to counteract gravity, allowing collapse to proceed. 

\begin{figure*}[htb!]
\epsscale{1.1}
\plotone{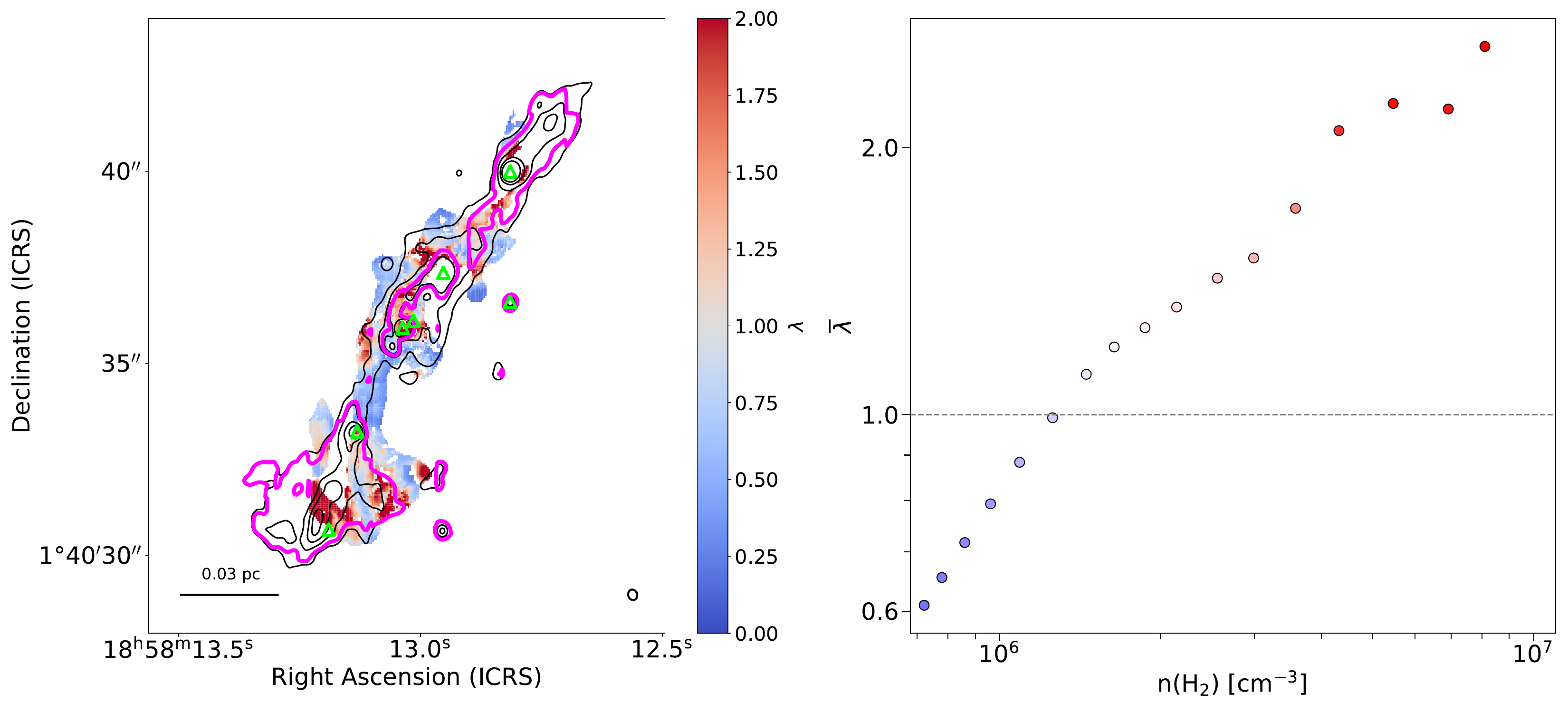}
\caption{(Left) The map of mass-to-flux ratio in G35. Black contours are the same as shown in Figure \ref{fig:poldisp}. The green triangles are the same as shown in Figure \ref{fig:magori}. (Right) Mean mass-to-flux ratio, $\overline{\lambda}$, as function of number density cut. The mean mass-to-flux ratio is calculated using mean column density and magnetic field strength within the number density cut. When $\overline{\lambda}$ is 1, the number density is 1.2 $\times$ 10$^6$ cm$^{-3}$, which is shown as the magenta contour in the left panel. 
\label{fig:mtob}}
\end{figure*}


Figure \ref{fig:mtob} shows the distribution of the mass-to-flux ratio in units of the critical value in G35. The reddish regions indicate that the observed values exceed the critical value. Most of them are located near the cores identified by \citet{Zhang2022}, which are shown as green triangles. The bluish regions indicate where the observed values are smaller than the critical value. The mass-to-flux ratios in G35 range from 0.1 to 6.0, with a mean value of 1.1 $\pm$ 0.8. The uncertainty is estimated through the error propagation applied to equation (\ref{eq:mass_flux_ratio}). 
It is clear that near the cores the gas is magnetically supercritical, in which the magnetic field cannot prevent gravitational collapse. We also note that the local estimate of the mass-to-flux ratio indicates whether, in the line of sight, there is enough mass at a pixel position to collapse in this position or not. However, the gas in each pixel could be gravitationally bound to a nearby core. We also note that regions showing large $\lambda$ values near the boundary may be affected by having a small number of segments compared to the central region, as well as by missing information at the edges introduced by our moving box method. 

We estimate magnetic criticality using the mean mass-to-flux ratio ($\overline{\lambda}$) within regions defined by number density cuts (Figure \ref{fig:mtob}). Within each number density cut, we calculate the mean column density and magnetic field strength. We obtain the mean mass-to-flux ratio by substituting these mean values into the equation (\ref{eq:mass_flux_ratio}). As the number density increases, magnetic criticality transitions from subcritical to supercritical. When $\overline{\lambda}$ = 1, the number density is 1.2 $\times$ 10$^6$ cm$^{-3}$, which is shown as a magenta contour in the left panel of Figure \ref{fig:mtob}.


%



\subsection{Magnetic Field Orientations at $\sim$0.1 pc Scales}

\begin{figure*}[htb!]
\epsscale{0.8}
\plotone{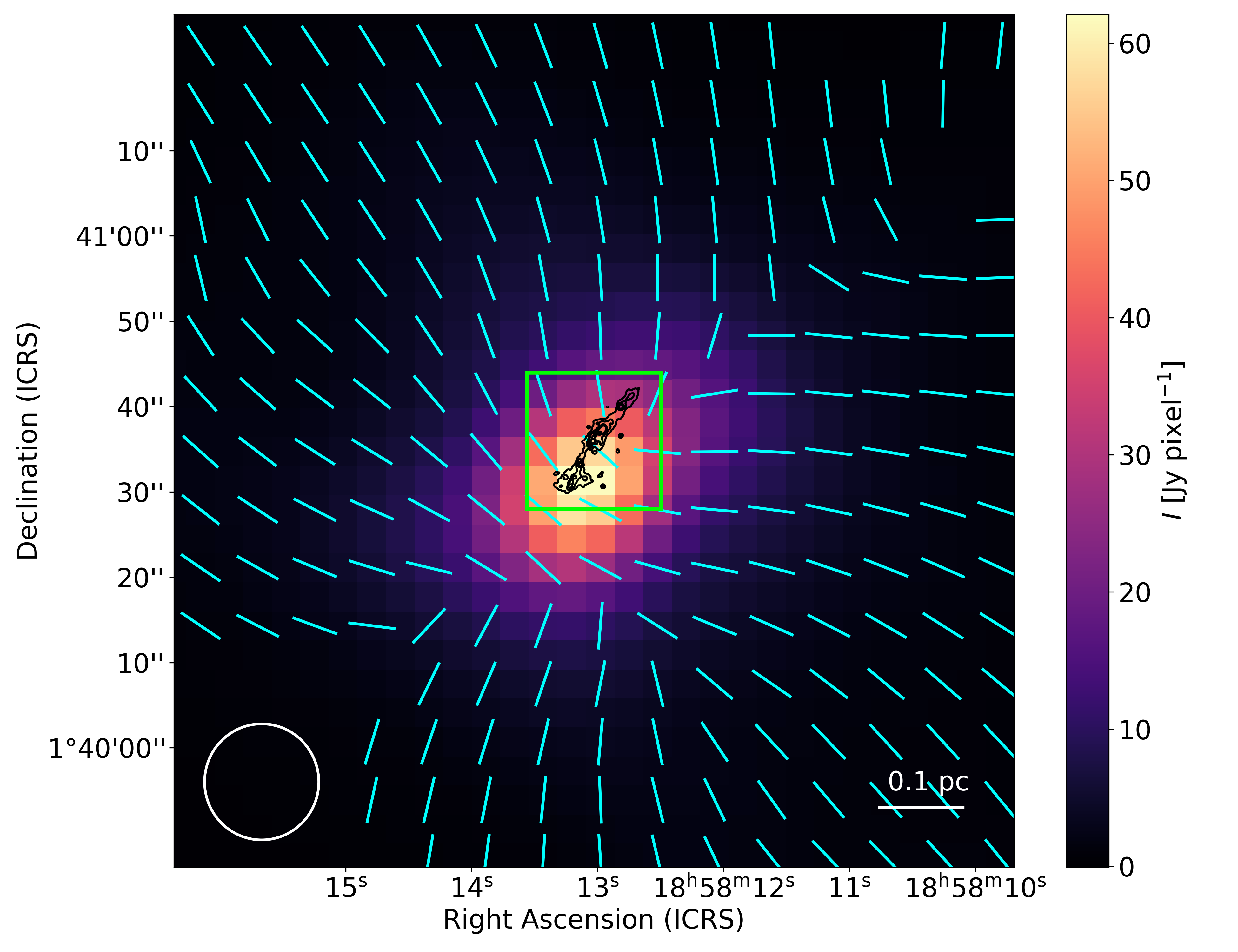}
\caption{The map of Stokes $I$ overlaid with magnetic field orientations (cyan segments) obtained with SOFIA at 154 $\mu$m. Black contours are the same as shown in Figure \ref{fig:poldisp}. Green box shows the area of the ALMA maps in preceding figures. The white circle indicates the beam size of SOFIA at 154 $\mu$m of 13\farcs6.
\label{fig:sofia}}
\end{figure*}

On a larger scale of $\sim$0.1 pc, compared to that of ALMA ($\sim$0.003 pc), the magnetic field orientation in G35 has been obtained using SOFIA at 154 $\mu$m with sub-pc resolution (Figure \ref{fig:sofia}). The background image in the figure shows the SOFIA Stokes $I$ data. The black contours indicate the dust continuum of G35 obtained using the ALMA observations at band 6. The beam size of the SOFIA data at 154 $\mu$m is 14\farcs0, a physical scale of $\sim$0.14 pc at a distance of 2.19 kpc. The FWHM of the SOFIA beam is about 45 times larger than that of the ALMA observations. The magnetic field structure obtained with SOFIA resembles an hourglass morphology toward the central dense region of G35 and is perpendicular to the filamentary structure revealed by ALMA. This sub-pc scale magnetic field morphology could be caused by converging gas flows toward the filament discussed in Section \ref{sec:velgrad}. The mean field orientation within the green box in Figure \ref{fig:sofia} is about 29 degrees measured from North to East. The size of the box is the same as in Figure \ref{fig:mag}. The circular mean magnetic field orientation obtained with ALMA within the lowest contour level shown in Figure \ref{fig:fila} is about -18 degrees. Even after performing the beam convolution and deriving the mean magnetic field orientation, the result remains consistent with the previously obtained value. However, the 12-m ALMA observations has missing flux and their maximum recoverable scale is 4.3$''$.

A similar transition has been reported in Serpens South by \citet{Pillai2020}, where the field changes from perpendicular to parallel to the filament above a visual extinction of 21 mag (H$_2$ column density $\sim$2.0 $\times$ 10$^{22}$ cm$^{-2}$). This threshold is comparable to our lowest contour level (1.4 $\times$ 10$^{22}$ cm$^{-2}$), supporting an interpretation that involves gas flows and gravitational collapse, consistent with our suggestions. \citet{Beuther2020} also showed that the magnetic field and gas are aligned parallel to the filament in the massive star-forming region G327.3. Simulations of filamentary molecular clouds undergoing global gravitational collapse suggested that magnetic field orientations can change from being perpendicular to parallel with respect to the filament, producing a characteristic U-shaped pattern \citep{Gomez2018}. We also found similar U-shaped magnetic field orientations at the northeastern edge of G35 (Figure \ref{fig:mag}). These previous studies support that gravity plays a dominant role in driving changes in the magnetic field orientation in G35. 

However, the discrepancy in magnetic field orientations between ALMA and SOFIA could be based on their observing frequencies.
 \citet{Fanciullo2022} reported that magnetic field orientations obtained using SOFIA at 154 and 214 $\micron$ and JCMT at 850 $\micron$ are perpendicular at the central dense region of Orion B. They explained that this angle difference could occur by dust alignment due to the anisotropic radiation \citep{Lazarian2007}, polarized thermal emission produced by anisotropic radiation \citep{Onaka1995}, or self-scattering by large grains \citep[e.g.,][]{Kataoka2015}. To clarify the change of the magnetic field orientations between different scales, we need intermediate-scale observations at a similar wavelength.


\subsection{Velocity Gradient}\label{sec:velgrad}

\begin{figure*}[htb!]
\epsscale{0.9}
\plotone{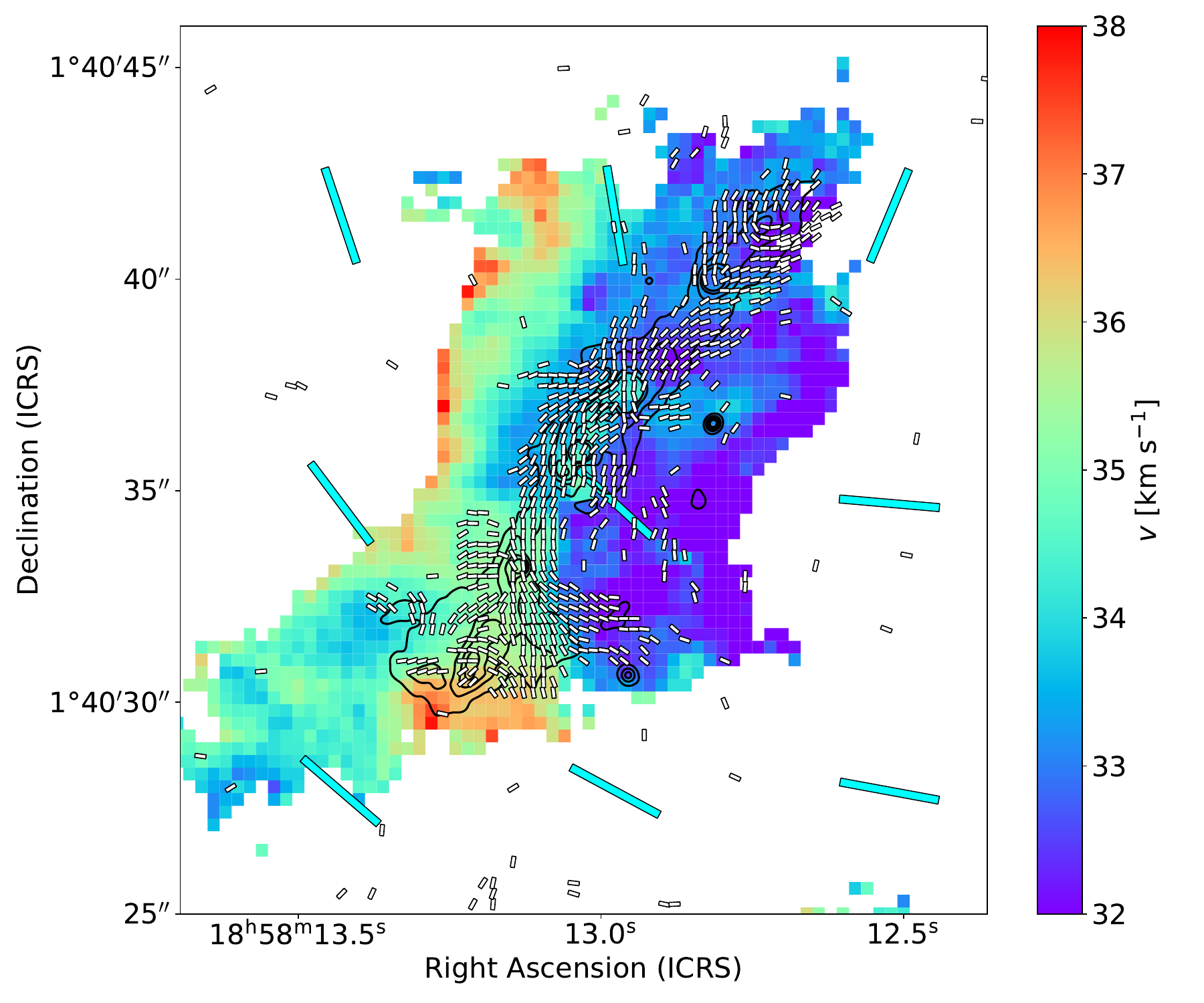}
\caption{The map of intensity-weighted mean velocity (moment 1) of H$^{13}$CO$^+$ obtained with ALMA, which is calculated within a velocity range from 30 to 40 km s$^{-1}$. White and cyan segments indicate magnetic field orientations obtained with ALMA and SOFIA, respectively. Black contours are the same as Figure \ref{fig:poldisp}.
\label{fig:mom1}}
\end{figure*}

\begin{figure*}[htb!]
\epsscale{1.1}
\plotone{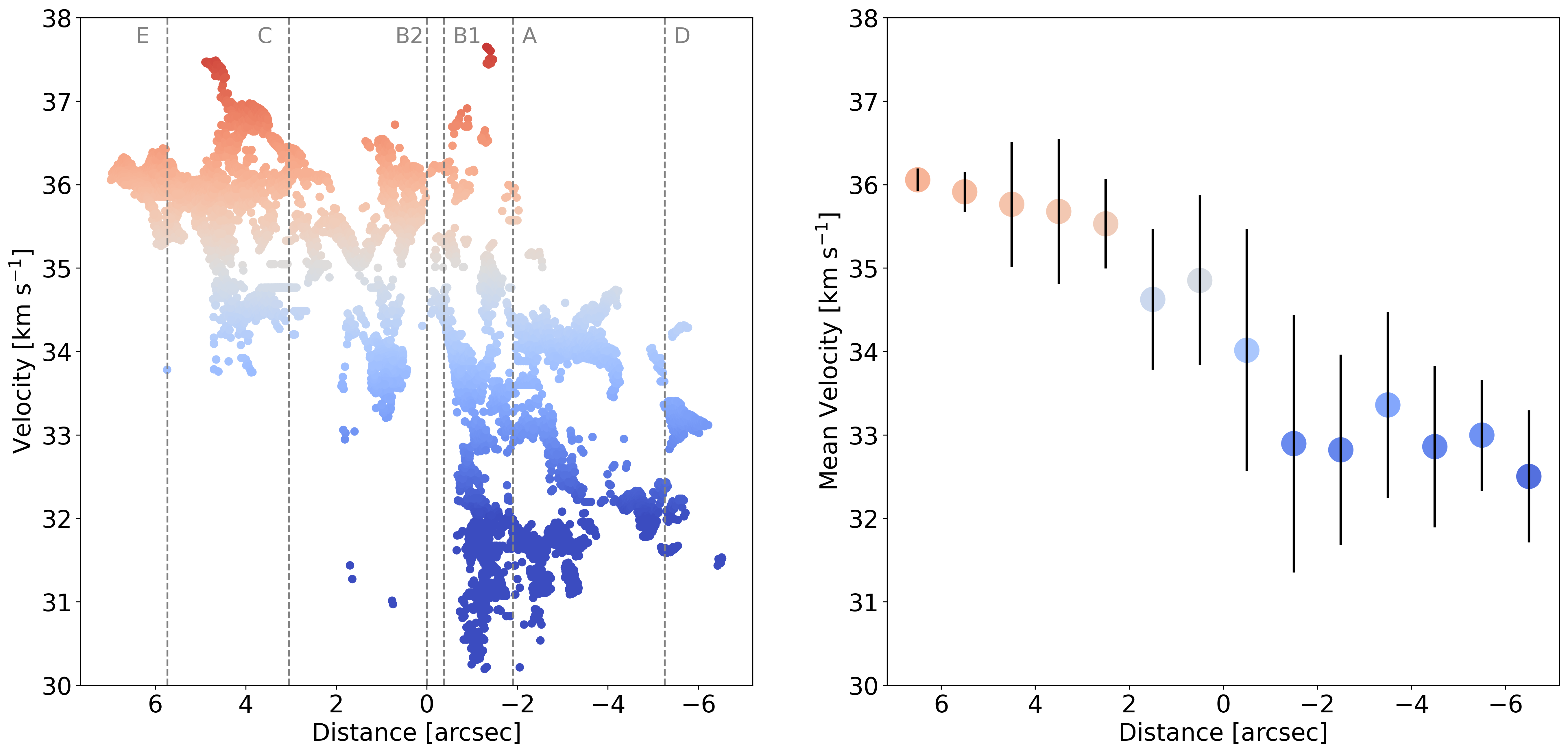}
\caption{(Left) The centroid velocities of H$^{13}$CO$^+$ for the velocity-coherent gas components described in section \ref{sec:veldisp} as a function of the distance from the B2 core of the central binary system in the G35 filament shown in Figure \ref{fig:magori}. The negative and positive distances mean eastern and western side from the core. The dashed lines indicate other core positions. The colors are shown red- and blue-shifed velocities compared to the LSR velocity of 35 km s$^{-1}$. (Right) The mean centroid velocities as a function of distance. The standard deviation is shown as an error bar. 
\label{fig:velgrad}}
\end{figure*}

To understand the change of magnetic field orientations from sub-pc to 0.003 pc scales, we investigate the velocity structure using H$^{13}$CO$^+$ data. To show the correlation between the gas flow and the magnetic field, we compare the field direction and the velocity gradient of the gas as traced by the H$^{13}$CO$^+$ spectral line. The intensity-weighted mean velocity (moment 1) map of H$^{13}$CO$^+$ shows a large-scale velocity gradient perpendicular to the filament (Figure \ref{fig:mom1}), which is roughly consistent with the mean magnetic field orientation of the SOFIA data. 
The large-scale magnetic field and velocity gradient could have initially contributed to the formation of the G35 filament. In the dense gas within the filament, no velocity gradients are evident simply by checking the moment 1 map due to the multiple velocity components.

We obtain the velocity gradient along the dense filamentary structure from the velocity-coherent gas component determined by the fof algorithm described in Section \ref{sec:veldisp}.
The left panel of Figure \ref{fig:velgrad} shows centroid velocities of the velocity-coherent gas structure as a function of the distance from the B2 core of the central binary system (B1 and B2). Figure \ref{fig:velgrad} shows a velocity gradient along the filament. The gray dashed lines show the positions of the six cores. The velocities are highly scattered in the nearby binary core, which could be caused by gravitational potential dragging matter toward the dense cores, or by outflow contamination. The eastern and western parts of the central binary system along the filament show red- and blue-shifted centroid velocities compared to the LSR velocity of 35 km s$^{-1}$ \citep[e.g.,][]{Jackson2006, Roman2009}. The red- and blue-shifted velocity components appear to be moving toward the center of the filament. Another possibility is that two gas flows having different velocities are converging together. Although velocity components are locally variable, we can find a global velocity gradient along the filament in the right panel of Figure \ref{fig:velgrad}. The mean velocity as a function of distance shows a velocity gradient of 29$\pm$3 km s$^{-1}$ pc$^{-1}$ (Figure \ref{fig:velgrad}). \citet{Qiu2013} showed a velocity gradient of H$^{13}$CO$^+$ (4--3) line obtained using SMA along the filament structure. They also found velocity gradients using CH$_3$OH and HC$_3$N lines, which is the rotation feature between cores A and B1/B2.

The usual interpretation of velocity gradients along filaments and streamers is that the gas is flowing towards the center of the gravitational potential of the clouds \citep[e.g.,][]{Kirk2013,Peretto2014,Olguin2023,Sanhueza2021,Sanhueza2025,Morii2025}. We therefore suggest that the infall along the filament is dragging magnetic field lines that otherwise would be perpendicular to the filament, as traced by the SOFIA observations. We follow the approach described in \citet{Kirk2013} to estimate the mass infall rate as $\dot{M}$ = $M \Delta V$/$\tan$($i$), in which $M = \mu m_\mathrm{H} N(\mathrm{H}_2) A$ is the mass of the filament, $A$ is the area of filament within the lowest contour level in Figure \ref{fig:fila}, $\Delta V$ is the velocity gradient of H$^{13}$CO$^+$ along the filament, and $i$ is the inclination angle of the filament. We estimate a mass infall rate of 0.9--3.4  $\times$ 10$^{-5}$ M$_\odot$ yr$^{-1}$ using uncertainty ranges and assuming an inclination angle of 30 and 60 degrees, respectively. This mass infall rate is comparable to those obtained in previous MagMaR targets within a factor of a few \citep[e.g.][]{Sanhueza2021, Sanhueza2025}.

\subsection{Magnetic Field and Core Fragmentation} \label{sec:magcore}

The six cores (A--E in Figure \ref{fig:magori}) along the filament in G35 have nearest-neighbor spacings ranging from 0.02 to 0.04 pc, measured as the shortest projected distance from each core to its closest-neighbor core.
We exclude the spacing between the central binary system (B1 and B2) of 0.004 pc. Linear fragmentation simulations assuming an infinite cylindrical filament show that the core separation is four times the filament width \citep[e.g.,][]{Inutsuka1992, Fischera2012, Nagasawa1987}. The mean filamentary width of G35 is about 0.015 pc, so the core separation predicted by the theory should be 0.06 pc. However, the core separation in the plane of the sky is less than 0.06 pc. If the inclination angle of the filament with respect to the plane of the sky is in the range of 48--71 degrees, the core spacing would be 0.06 pc. However, it is hard to constrain the inclination angle of the filament. Previous observational studies have also reported core spacings comparable to filament widths or less than four times the width \citep[e.g.,][]{Tafalla2015, Zhang2020, Shimajiri2023, Pineda2023}, which is consistent with our results. Some models include a magnetic field perpendicular to the filament, in which the core spacing is slightly less than four times the filament width \citep{Hanawa2015, Hanawa2017, Hanawa2019, Seifried2015}. The fragmentation of filaments called the sausage instability \citep[e.g.,][]{Contreras2016} assumes an isothermal cylinder model with a helicoidal magnetic field \citep{Nakamura1993}. Assuming this scenario, we obtain a core spacing of 0.09 pc when the central density is 1.5 $\times$ 10$^6$ cm$^{-3}$ from the peak intensity of the averaged profile in Appendix \ref{sec:fwhm}, and assuming a central magnetic field strength equal to the peak magnetic field strength of 4.4 mG. This core separation is larger than that obtained by the filament fragmentation model, so we conclude that the filament fragmentation model makes a better prediction on the core separation than does the sausage instability model.  

Based on the velocity gradient, we estimate the timescale for core migration along the filament,
\begin{equation} 
\Delta t = \frac{\nabla d}{\nabla v \nabla l\cos i },
\label{eq:time}
\end{equation}
where $\nabla d$ is the distance moved from the initial to the current core position, $\nabla v$ is the velocity gradient along the filament (29 km s$^{-1}$ pc$^{-1}$), $\nabla l$ is the initial separation between cores, and $i$ is the inclination angle. We assume $i = 45^\circ$. The estimated timescale for a core to move 0.02 pc is $\sim$2.5 $\times 10$$^4$ years. The typical timescale for massive star formation is $\sim$10$^5$ years \citep[e.g.,][]{McKee2002, Sabatini2021}. Given this timescale, the observed core spacing may result from the migration of cores following filament fragmentation.

\subsection{Schematic View}

\begin{figure*}[htb!]
\epsscale{1.1}
\plotone{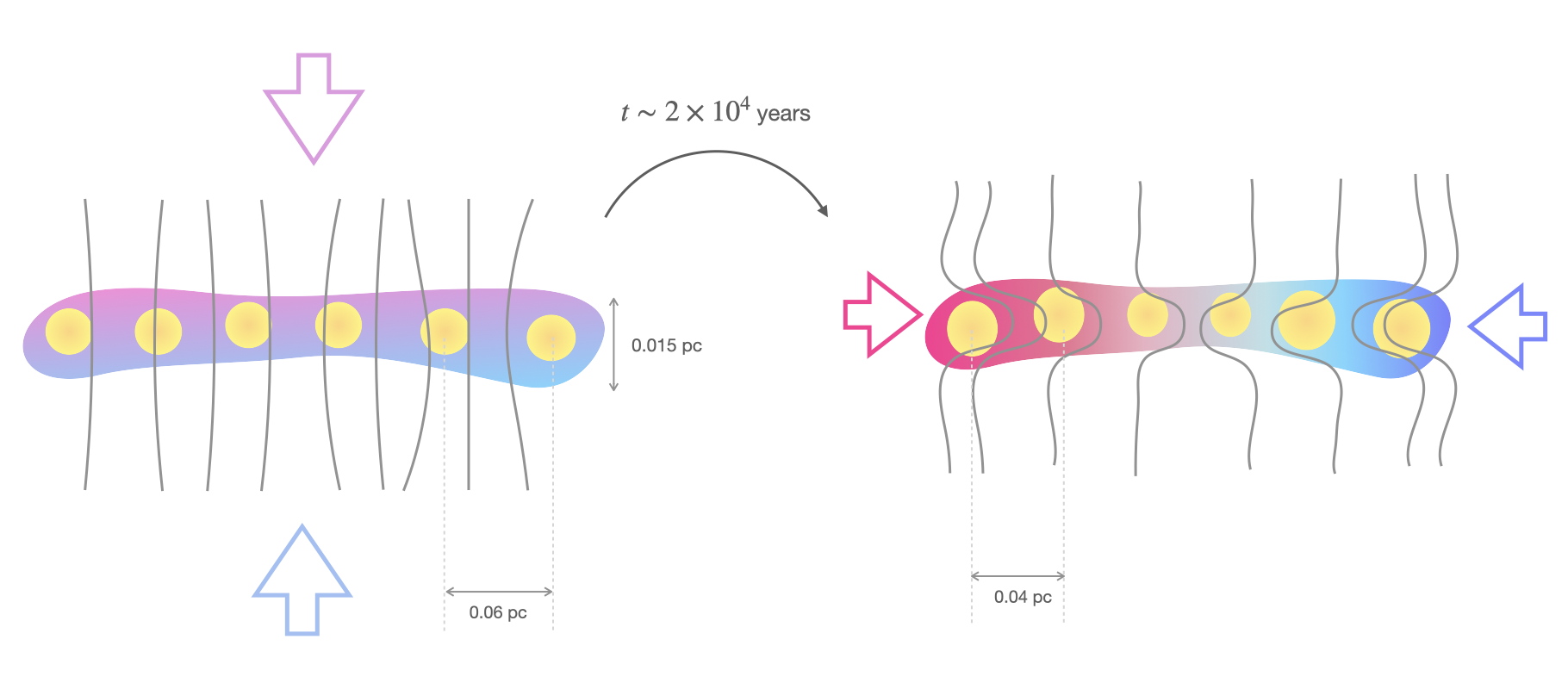}
\caption{The schematic view of G35. The large-elongated structure and yellow circles indicate the filament and dense cores in G35. The grey lines show magnetic field lines. Magnetic field lines are initially perpendicular to the major axis of the filament (Left). 
As the density increases, the gas moves and drags the magnetic field lines
(Right). Cores are also moved along the filament making short separation. The black arrow means the time evolution of the filament. Colored arrows show gas movement based on velocity gradients.
\label{fig:view}}
\end{figure*}

Based on our analysis, we suggest a schematic view of the formation of G35 (Figure \ref{fig:view}). Initially, low-density gas moves along magnetic field lines and forms a filamentary structure. The filament is located at the central part of a large hourglass-like large-scale magnetic field structure. The initial filament could have been made by material guided by the large-scale magnetic field. The initial magnetic field orientations are revealed by SOFIA, perpendicular to the filament, as shown to the left of Figure \ref{fig:view}. The intensity-weighted mean velocity map shows the velocity gradient of the lower-density gas perpendicular to the major axis of the filament. The cores in the filament could have been formed in local high-density regions by fragmentation with a separation of four times the filament width, 0.06 pc.

 As time goes on, the right side of Figure \ref{fig:view} shows the magnetic field dragged along the filament. 
As density increases, gas flow along the filament changes the direction of the magnetic field from perpendicular to parallel to the filament. The curved magnetic field at the north-west tip of the filament is likely the result of gas flow along the filament. Magnetic field lines are aligned along the filament with a mean angle difference of 29 degrees. When considering the velocity gradient, the cores can move 0.01--0.02 pc along the filament within the time scales of the formation of high-mass stars. 


\section{Summary}\label{sec:Sum}

As part of the MagMaR survey, we present dust polarization and H$^{13}$CO$^+$ observations of G35 obtained using ALMA in Band 6. The dust continuum of G35 shows a filamentary structure in which the magnetic field orientation is aligned along the major axis of the structure.

$\bullet$ We estimate the dispersion of the polarization angle within a small moving box with a size of 3 $\times$ 3 beams. By moving the box in the whole region, we make a map of the polarization angle dispersion.

$\bullet$ We obtain the volume density map using the kinetic temperature map obtained from VLA observations of NH$_3$ and dust continuum emission ALMA observations. The velocity dispersion map was made using the H$^{13}$CO$^+$ observations. We conduct multiple Gaussian fitting toward each spectrum and identify the largest coherent velocity structure using the fof algorithm. 

$\bullet$ Using the maps of polarization angle dispersion, volume density, and velocity dispersion, we estimate the distribution of magnetic field strengths ranging from 0.2 to 4.4 mG with a mean value of 0.8 $\pm$ 0.4 mG.

$\bullet$ To investigate the relative importance between the magnetic field and gravity, we estimate mass-to-flux ratios toward G35 varying from 0.1 to 6.0 with a mean value of 1.1 $\pm$ 0.8. The dense regions of G35 are magnetically supercritical, indicating that the magnetic field is not strong enough to support the cores against gravitational collapse. 

$\bullet$ Six dense cores are fragmented along the long axis of the filament in G35, along which a H$^{13}$CO$^+$ velocity gradient of 27 km s$^{-1}$ pc$^{-1}$ is detected. 


$\bullet$ The large-scale magnetic field lines obtained using SOFIA shows an hourglass morphology converging toward the central dense region observed using ALMA. The segments obtained by SOFIA seem to be connected to segments in localized regions on the filament. The mean SOFIA magnetic field direction is also aligned at large scales with the velocity gradient of the lower density gas, as shown in the moment 1 map of H$^{13}$CO$^+$. 

$\bullet$ Based on our results, we suggest a formation and evolution scenario for G35. Magnetic field lines would initially regulate the formation of filaments based on the preferential movement along field lines. As the density increases, increasing the gravitational force, gas flows along the filament and drags the magnetic field.

\software{CASA \citep{CASA2022}}
\facilities{ALMA, SOFIA}

\begin{acknowledgments}
Data analysis was in part carried out on the Multi-wavelength Data Analysis System operated by the Astronomy Data Center
(ADC), National Astronomical Observatory of Japan. This paper makes use of the following ALMA data: ADS/JAO.ALMA\#2017.1.00101.S and ADS/JAO.ALMA\#2018.1.00105.S. ALMA is a partnership of the ESO (representing its member states), NSF (USA) and NINS (Japan), together with NRC (Canada), MOST and ASIAA (Taiwan), and KASI (Republic of Korea), in cooperation with the Republic of Chile. The Joint ALMA Observatory is operated by ESO, AUI/NRAO, and NAOJ.
We thank Dr. Chao Wang and Ke Wang for sharing the NH$_3$ multi-transition fitting results from JVLA. 
 PS was partially supported by a Grant-in-Aid for Scientific Research (KAKENHI Number JP23H01221) of JSPS.
J.M.G. acknowledges support from the PID2023-146675NB grant funded by MCIN/AEI/10.13039/501100011033, and by the program Unidad de Excelencia Mar\'{\i}a de Maeztu CEX2020-001058-M.
I.W.S. acknowledges financial support for this work provided by NASA through the award \#09\_0016 issued by USRA.
M.T.B. acknowledges financial support through the INAF Large Grant {\it The role of MAGnetic fields in MAssive star formation} (MAGMA).
J.L. is partially supported by Grant-in-Aid for Scientific Research (KAKENHI Number 23H01221 and 25K17445) of the Japan Society for the Promotion of Science (JSPS). 
P.S. was partially supported by a Grant-in-Aid for Scientific Research (KAKENHI No JP24K17100) of the Japan Society for the Promotion of Science (JSPS).
L.A.Z. acknowledges financial support from CONACyT-280775, UNAM-PAPIIT IN110618, and IN112323 grants, M\'{e}xico.
Y.C. was partially supported by a Grant-in-Aid for Scientific Research (KAKENHI  number JP24K17103) of the JSPS. 
E.C. acknowledges the support from Core Research Grant (CRG; sanction order number CRG/2023/008710) awarded by Anusandhan National Research Foundation (ANRF) under Science and Engineering Research Board (SERB), Govt. of India.
K.P. is a Royal Society University Research Fellow, supported by grant number URF\textbackslash{}R1\textbackslash{}211322.
\end{acknowledgments}

\appendix

\section{Filament width}\label{sec:fwhm}

\begin{figure*}[htb!]
\epsscale{0.8}
\plotone{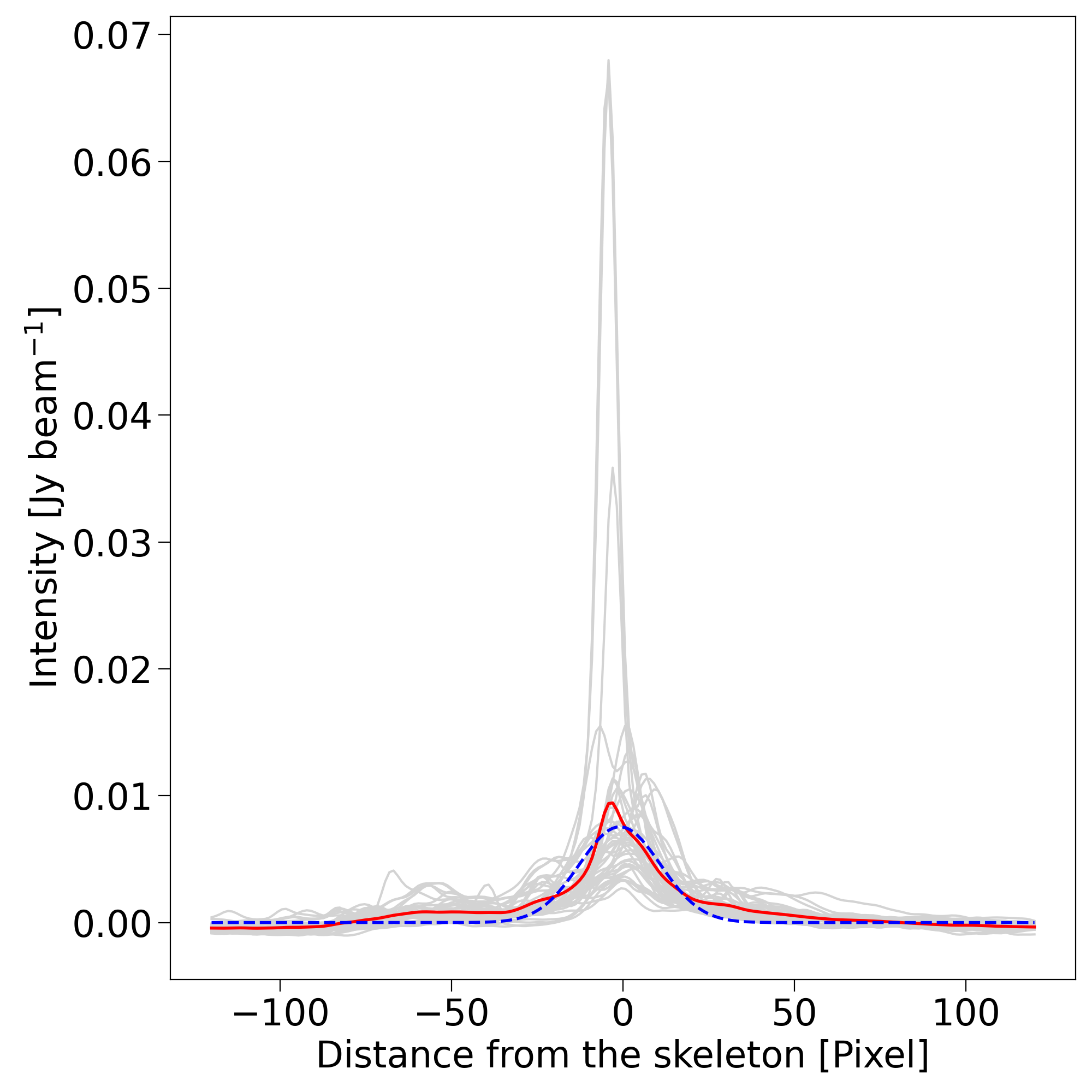}
\caption{Intensity of dust continuum emission across the filaments. The horizontal axis shows the distance from the skeleton shown in the left panel Figure \ref{fig:fila}. The gray lines show the intensity profiles perpendicular in each pixel on the skeleton with a spacing of a beam size. The red line shows the averaged value of the gray lines. The blue dashed line is obtained by fitting the red line with a single Gaussian. 
\label{fig:fwhm}}
\end{figure*}

To determine the filament width, we analyze intensity profiles perpendicular to the skeleton orientation. The skeleton orientation is determined as the circular mean of angles averaged over each pixel and its neighboring pixels within one beam area. We estimate the intensity along the direction perpendicular to the skeleton. The gray lines in Figure \ref{fig:fwhm} show individual intensity cuts taken along the skeleton with a spacing of one beam size. The red line represents the average of these gray profiles. We fit the red line with a single Gaussian function. The FWHM of the Gaussian fit is 0.015 pc, which is adopted as the width of the filament.

\end{document}